\documentclass[english,twocolumn]{revtex4-1}
\usepackage[T1]{fontenc}
\usepackage[latin1]{inputenc}
\usepackage{babel}
\usepackage{epsfig}
\usepackage{amsmath}
\usepackage{amssymb}
\usepackage{epstopdf}
\usepackage{overpic}
\usepackage{color}

\usepackage{natbib}
\usepackage{comment}
\begin{document}

\title
{\bf The Persistence of Uphill Anomalous Transport in Inhomogeneous Media}
\author{C. Mulhern*}
\footnotetext{The text of the footnote goes here.}
\medskip
\medskip
\medskip
\affiliation{Max Planck Institute for the Physics of Complex System, 01187 Dresden, Germany}

\email[Corresponding author: ]{mulhern@pks.mpg.de}

\begin{abstract}
\noindent 
For systems out of equilibrium and subjected to a static bias force it can often
be expected
that particle transport will usually follow the direction of this bias. However,
counter-examples exist
where particles exhibit uphill motion (known as \emph{absolute negative mobility} - ANM), particularly in the
case
of
coupled particles. Examples in single particle deterministic systems are less common. Recently, in one such
example, uphill
motion was shown to occur for an inertial driven and damped particle in a spatially symmetric periodic
potential. The source of this anomalous transport was a combination of two periodic driving signals which 
together are asymmetric under time reversal. In this paper we investigate the phenomena of ANM for a
deterministic particle evolving in a periodic and symmetric potential subjected to an external
unbiased periodic driving and nonuniform \emph{space-dependent} damping. It will be shown that this system
exhibits a complicated response behaviour as certain control parameters are varied, most notably being,
enhanced parameter regimes exhibiting ANM as the static bias force is increased.
Moreover, the solutions exhibiting ANM are shown to be, at least over intermediate time periods,
superdiffusive, in contrast to the solutions that follow the bias where the diffusion is normal.
\end{abstract}

\maketitle
\section{Introduction}
\noindent
The dynamics of systems modelling the evolution of single driven and damped particles continues to be of
interest. One reason is the rich behaviour present in such models. Another is that these relatively simple
systems allow for the analysis and observation of real physical phenomena with only minimal
resources. In particular, the transport of particles in symmetric and periodic potential landscapes has
attracted considerable interest ~\cite{Acevedo1, Reimann1}. Such potentials lend themselves to a vast number
of applications including Josephson junctions \cite{Majer1}, charge density waves,
nanoengines \cite{Astumian1}, and transport in biological systems \cite{Hanggi1}.

The prototypical equation for such models takes the form 

\begin{equation}\label{eq:proto}
 \ddot{q} = -\gamma\dot{q} - V'(q) + F(t)
\end{equation}

\noindent
where $\gamma$ is the damping parameter, $V(q)$ is the system potential, and $F(t)$ is a time-dependent 
damping; $V$ and $F$ are both usually bounded and periodic. The dot and prime denote differentiation with
respect to time $t$, and coordinate $q$ respectively. The symmetry properties of these models are now
well understood \cite{Flach1,Denisov5}. In short, if
the system potential and the external driving satisfy certain spatial and temporal symmetries, then each
trajectory will have a counterpart whose velocity is of the same magnitude, but of different sign. This has
important consequences for the net flow (often called the \emph{current} and is defined more precisely later)
as, if each trajectory has a
counterpart whose velocity differs only by a change of sign, then the net flow will be zero. Thus, a number of
studies have investigated the effects, with respect to the net flow, when these symmetries are broken. For
example, numerous studies have considered the driven and damped dynamics of a particle evolving in a
periodic, but asymmetric potential \cite{Mateos1, Mateos2, Kenfack1, Vincent1}. They observed a nonlinear
response behaviour to changes in the driving amplitude, including multiple current reversals. In the
Hamiltonian limit $\gamma=0$, the focus has been on how these asymmetries influence the sticking episodes to
regular
transport supporting islands in the chaotic part of phase-space \cite{Schanz1, Schanz2, Denisov2}. Obtaining
directed particle transport in systems with zero-average forces has become known as the \emph{ratchet effect}
\cite{Denisov8}.

Recently, \cite{Saikia1} considered an alternative to the more common spatially
uniform damping. They studied a system of the form given by Eq.~\ref{eq:proto}, with symmetric potential and
driving, where the constant coefficient of friction $\gamma$ was replaced by a space
dependent term. They found that the frictional inhomogeneity mimics the role played
by asymmetric potentials and/or external driving forces, resulting in non-zero
net flow, i.e. the ratchet effect. Motivations for such studies come from a variety of sources. For
example, in Josephson Junctions a phase-dependent damping can represent an interference term between the pair
and quasiparticle currents \cite{Leppakangas1, Falco1} (in the latter the authors also give a thorough
phase-space analysis of
such a
junction).

An interesting extension to problems with an externally modulated potential comes when a dc-bias is
introduced, serving as a constant tilt to the potential landscape ~\cite{Alatriste1, Borromeo1, Eichhorn2,
Speer2, Speer1, Machura1, Eichhorn1}. These studies have examined the fascinating phenomena of \emph{absolute
negative mobility} (ANM)
where a particle can travel in the direction opposite to a constant applied force. Most of the studies so far
have looked at noise induced ANM. Further, it was proven that for the over-damped dynamics
of Brownian particles, where inertial effects are negligible, the solutions may not exhibit ANM
\cite{Speer2}.

Studies of these inherently biased systems are important as they find application in a number of areas. For
example, in the transport of biomolecules where the separation of particles
may be desirable \cite{ZuRuiLi1}, this separation becomes inherently difficult when the particles are working
against an additional load. Therefore, finding regimes where particles move against an applied load becomes
extremely important. Further, ANM has recently been observed experimentally in the domain of Josephson
Junctions where the
related phenomena is known as \emph{Negative Absolute Resistance} \cite{Nagel1}. The authors were able verify
theoretical predictions obtained from a model of a damped Brownian particle in
one dimension \cite{Speer2,Speer1}.

Less common are works on the ANM phenomena in single particle deterministic systems, i.e. the noiseless case.
This has been detailed
in only a few studies, for example \cite{Alatriste1, Machura1, Speer2,Speer1}. A recent study
\cite{Du1}, in an attempt to mimic
the roll played by noise in previous works, considered a ``vibrational motor'' -- a system where additional
driving terms yield stochastic-like (yet deterministic) dynamics. ANM was observed in this system in
regimes where it was solely induced by noise (when the additional driving terms are absent).

To the authors' knowledge, the effect of absolute negative mobility has not been observed in systems with a 
frictional inhomogeneity. Illustrating such an effect will be the focus of the present study. In particular,
we investigate the transport processes of single particles evolving in a symmetric and periodic potential,
subjected to an unbiased external ac-driving and a static dc-bias. It will be shown that a frictional
nonuniformity can induce the phenomena of absolute negative mobility. Moreover, the mechanism that allows for
such an astonishing effect is different from those presented to-date, and this
will also be discussed.

The paper is organised as follows. In the next section we outline the system under investigation, and discuss
some of its important properties. Here, the main observable of interest, i.e. \emph{particle current}, will
also be presented. Numerical results, pertaining to the particle current, will then be presented in
\S\ref{sec:current}. A discussion then follows in \S\ref{sec:anm} on the mechanism and phase space
structures that allow for the occurrence of ANM in the system under consideration. Further, the dynamics
will be characterised in terms of rates of diffusion. We finish with a summary of the results.

\section{System}
We study the dynamics of a driven and damped particle evolving in a symmetric 
and periodic ``washboard'' potential. The potential, in addition to the time
periodic modulations of its inclination, will also be subjected to a static
dc-bias force. Further the strength 
of the damping will be space dependent. The equation of motion for this system is given by 

\begin{equation}
 \ddot{q} = -\gamma(q)\dot{q} + A\cos(\omega t) + \cos(q) + F
\end{equation}

\noindent
where $q = q(t)$ represents the spatial coordinate of the particle at time $t$,
and with potential $V(q) = -\sin(q)$, and $\gamma(q) =
\gamma[1-\lambda\sin(q+\phi)]$, both of spatial period $L=2\pi$. The particle is
driven by
a zero average time-periodic driving force of
amplitude $A$ and frequency $\omega$, and the magnitude of the static bias force is represented by the
parameter $F$. In addition, the space dependent damping is regulated by three
parameters, namely $\gamma$,$\lambda$, and $\phi$, which control the maximal
amplitude of the damping coefficient, determine the systems
inhomogeneity, and determine the phase difference between the potential and the damping coefficient (which
are of the same period). 

As a physical realisation of such a system consider a resistively and capacitively shunted Josephson Junction
subjected to ac- and dc-currents. The corresponding equations of motion model the phase difference across the
junction. The phase dependent term (often called the ``$\cos\phi$'' term) accounts for interference between
the Cooper-pair and quasiparticle currents \cite{Leppakangas1, Falco1}. 
As mentioned in the introduction, ANM has been observed in a Josephson Junction experimentally \cite{Nagel1}.

It is worth examining the symmetry properties of this system for the special cases related to $F=0$. These
properties determine whether or not a current (defined in this section) can emerge in the ensemble dynamics.
As stated in \cite{Flach1},
the breaking of each system symmetry is required before a current can emerge. Consider these special
cases: firstly $F=0$ and $\lambda=0$, and secondly $F=0$, $\lambda \neq 0$ and $\phi = n\pi$ ($n \in
\mathbb{Z}$). In both cases the transformation $q\rightarrow -q+\pi$, $t\rightarrow t+T/2$ ($T=2\pi/\omega$)
yields a new
trajectory with average velocity which differs from that of the original only by sign. Thus a zero current
results.
This transformation for $\lambda\neq 0$ and $\phi \neq n\pi$ ($n \in \mathbb{Z}$) does not necessarily
produce counter-propagating trajectories and thus the necessary conditions for the ratchet effect to occur
have been
created. The dynamical effects of $\phi \neq n\pi$ ($n \in \mathbb{Z}$) are now discussed.

\begin{figure}
\includegraphics[height=4.8cm, width=5.8cm]{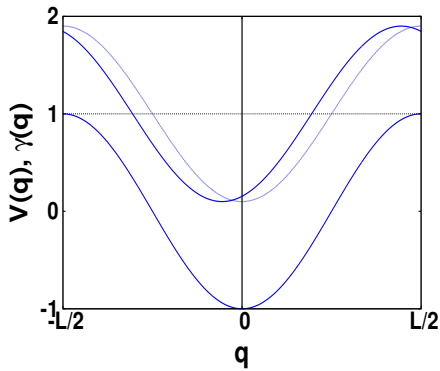}
\caption{Shown, over one spatial period $L$, are the potential $V(q)$ (bottom curve), and the
corresponding nonuniform coefficient of friction $\gamma(q)$ for two values of the phase $\phi$; the curve
with minimum centred at $q=0$ corresponds to phase $\phi=0$, while the curve with a minimum centred to the
left of $q=0$ results when $\phi=0.35$; here $\lambda=0.9$.}
\label{fig:bias}
\end{figure}

Following similar lines to the discussion of \cite{Reenbohn1}, we outline how the frictional nonuniformity can
be used to induce the ratchet effect in the absence of a dc-bias force. Fig.~\ref{fig:bias} shows the
potential $V(q)$, and the strength of the coefficient of damping (for two values of the phase $\phi$) over one
spatial period $L$. With $\phi = 0$ the damping coefficient is symmetric about the potential minimum with the
result that neither motion to the left, nor the right, is favoured with respect to the nonlinear damping term.
With a nonzero phase ($\phi=0.35$ in the this case) symmetry with respect the potential minimum is broken.
Looking at the curve corresponding to $\phi=0.35$ in Fig.~\ref{fig:bias} it can be seen that the damping
coefficient to the left of the potential minimum is, on average, smaller than
that to the right of the
potential minimum, thus favouring motion to the left. It is this mechanism that allows for the emergence of a
nonzero current \cite{Saikia1}. The ratchet effect, induced by this mechanism, will be exploited in the
present work.

To gain a quantitative perspective on how the frictional inhomogeneity parameter $\lambda$ influences the
dynamics we compute the current $\emph{v}$. That is, we calculate the time averaged mean velocity for an
ensemble of initial conditions, i.e.

\begin{equation}
\emph{v} = \dfrac{1}{T_{s}}\int_{0}^{T_{s}} dt\langle p(t)\rangle,
\end{equation}

\noindent
where $T_{s}$ is the simulation time, and the ensemble average is given by

\begin{align}
\langle p(t)\rangle = \dfrac{1}{N}\sum_{n=1}^{N}p_{n}(t),\label{eq:current}
\end{align}

\noindent
with $N$ being the number of initial conditions. Numerical results related to the current will be presented in
the next section.

\section{Current}\label{sec:current}
In this section we discuss the numerically computed current. The initial conditions have been chosen such
that the $q_n(0)$ are uniformly distributed in the potential well centred at the origin, with $p_n(0) = 0$ for
all $n \in N$. For computation of the long time average numerical integration is performed using a
fourth-order
Runge-Kutta method, over a simulation time interval $T_s = 10^5$ with step size $dt=0.01$. The ensemble
average is calculated using an ensemble of $N=1000$ initial conditions.

\begin{figure}
\includegraphics[height=4.8cm, width=5.8cm]{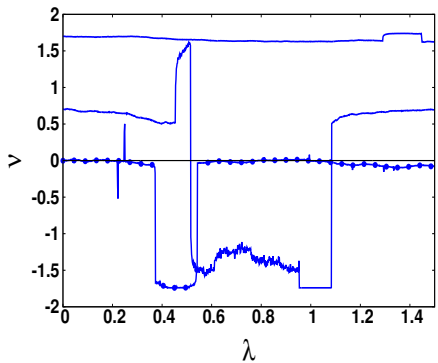}
\caption{The current, computed for three values of the static bias force $F$, as a function of $\lambda$.
The curve with dots corresponds to a dc-bias of $F=0.0$, the middle curve to $F=0.1$, and the upper curve
to $F=0.2$. The remaining parameters are $\gamma=0.108$, $\phi=0.35$, $A=1.512$, and $\omega=0.58$.}
\label{fig:current}
\end{figure}

Fig.~\ref{fig:current} shows the current, as a function of $\lambda$, for different values of the static bias
force $F$. For $F=0$ the current is in the main close to zero. However, there exists a window of
$\lambda$ values such that motion to the left is promoted ($0.37 \lesssim \lambda\lesssim  52$). The direction
of the current in this window is most certainly induced, not only by the choice of $\lambda$, but also by the
specific choice of the phase $\phi=0.35$
(see Fig.~\ref{fig:bias}). Moreover, another choice of $\phi$ can induce a
current that moves to
the right. Note, that motion to the left in this case does not qualify as negative
mobility as the particle is not working against an external load. This can only happen for $F\neq0$.
Increasing the bias force to $F=0.1$, we see that a window, of significant extent, opens which supports
ANM. Outside of this window the current follows the bias, i.e. there is a positive
current. Importantly, most of the $\lambda$ values inside this window of negative mobility produce a zero
current when the static bias force is switched off. Thus, one can conclude that this effect of negative
mobility is induced by the static bias force, rather than through a carefully chosen phase $\phi$, i.e
it is the tilting of the potential that results in uphill motion. The reason for the fluctuations of \emph{v}
in this window is due to the coexistence of attractors supporting transport in opposite directions (see
Fig.~\ref{fig:bifur}). However, for $0.95 \lesssim \lambda \lesssim 1.08$ there is single attractor in phase
space supporting uphill motion. This helps explain why the current remains constant within this window, and
further why the current has an increased magnitude. 

Such windows of absolute negative mobility exist for $F < F_{crit} \approx 0.17$ .
Remarkably, as $F$ is increased from $F=0$, the size of the window supporting ANM grows approaching almost
three times the $F=0$ size. However, this behaviour eventually ceases and as $F\rightarrow F_{crit}$ from
below, the windows become of smaller and smaller extent. Beyond $F_{crit}$,
solutions exhibiting negative mobility no longer exist, and instead follow the direction of the bias force,
resulting in a positive current. An example of this is shown for $F=0.2$ where the current is $\nu \approx
1.75$ for the entire range of $\lambda$.

To see that the occurrence of ANM in this system is indeed dependent on the 
frictional inhomogeneity, consider again the curve related to $F=0.1$ in Fig.~\ref{fig:current}. Starting from
zero frictional inhomogeneity ($\lambda=0$), the current is positive, i.e. the current is in the direction of
the bias. Upon increasing $\lambda$ this remains true until $\lambda\approx 0.51$ where there is an abrupt
change in the direction of the current. The current then remains negative with increasing $\lambda$ until a
second critical value $\lambda\approx 1.08$ where the current again becomes positive. Thus, for the constant
dc-bias $F=0.1$, ANM is possible only for certain values of the frictional inhomogeneity
parameter $\lambda$. 

In the next section, we will discuss the phase space structures that allow for such counter-intuitive
motion. Moreover, the mechanism that produces uphill motion will also be discussed.

\section{Absolute Negative Mobility}\label{sec:anm}
In this section it is our aim to gain further understanding of the phase-space structures that facilitate
this uphill motion, and to look more closely at the underlying mechanism that allows for negative mobility. 

\begin{figure}
\includegraphics[height=4.8cm, width=5.8cm]{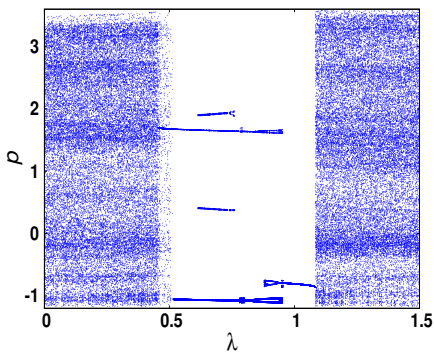}
\caption{Bifurcation diagram, as a function of $\lambda$, for $F=0.1$. The remaining parameters are given in
Fig.~\ref{fig:current}.}
\label{fig:bifur}
\end{figure}

Let us first consider the bifurcation diagram for the curve related to $F=0.1$ in Fig.~\ref{fig:current}. The
results, obtained by stroboscopically sampling trajectories after each period of the driving (omitting a
transient), are contained in Fig.~\ref{fig:bifur}. It can be seen that, for the range of $\lambda$ considered,
this system supports aperiodic chaotic solutions, and periodic solutions. With regard to the window of
ANM observed in Fig.~\ref{fig:current}, the corresponding window in Fig.~\ref{fig:bifur}
supports only periodic solutions. In contrast, it would appear that the solutions following the
direction of the dc-bias evolve chaotically. This behaviour would help explain why, for the majority of
$\lambda$ values (with $F=0.1$), the current in the window of ANM is of greater magnitude than
for the $\lambda$ values corresponding to a positive current. 

Just like for the current, the transition from chaotic motion to periodic motion is abrupt. Sharp
transitions from chaotic to periodic motion (and vice versa) related to the transition from downhill to uphill
motion have been
observed before in the case of coupled particle \cite{Mul2}. Moreover, the exact reasons behind
current reversals in general single particle systems of the form Eq.~\ref{eq:proto} remains open to debate
\cite{Mateos1, Barbi1, Kenfack1}. 

\begin{figure}
\includegraphics[height=4.8cm, width=5.8cm]{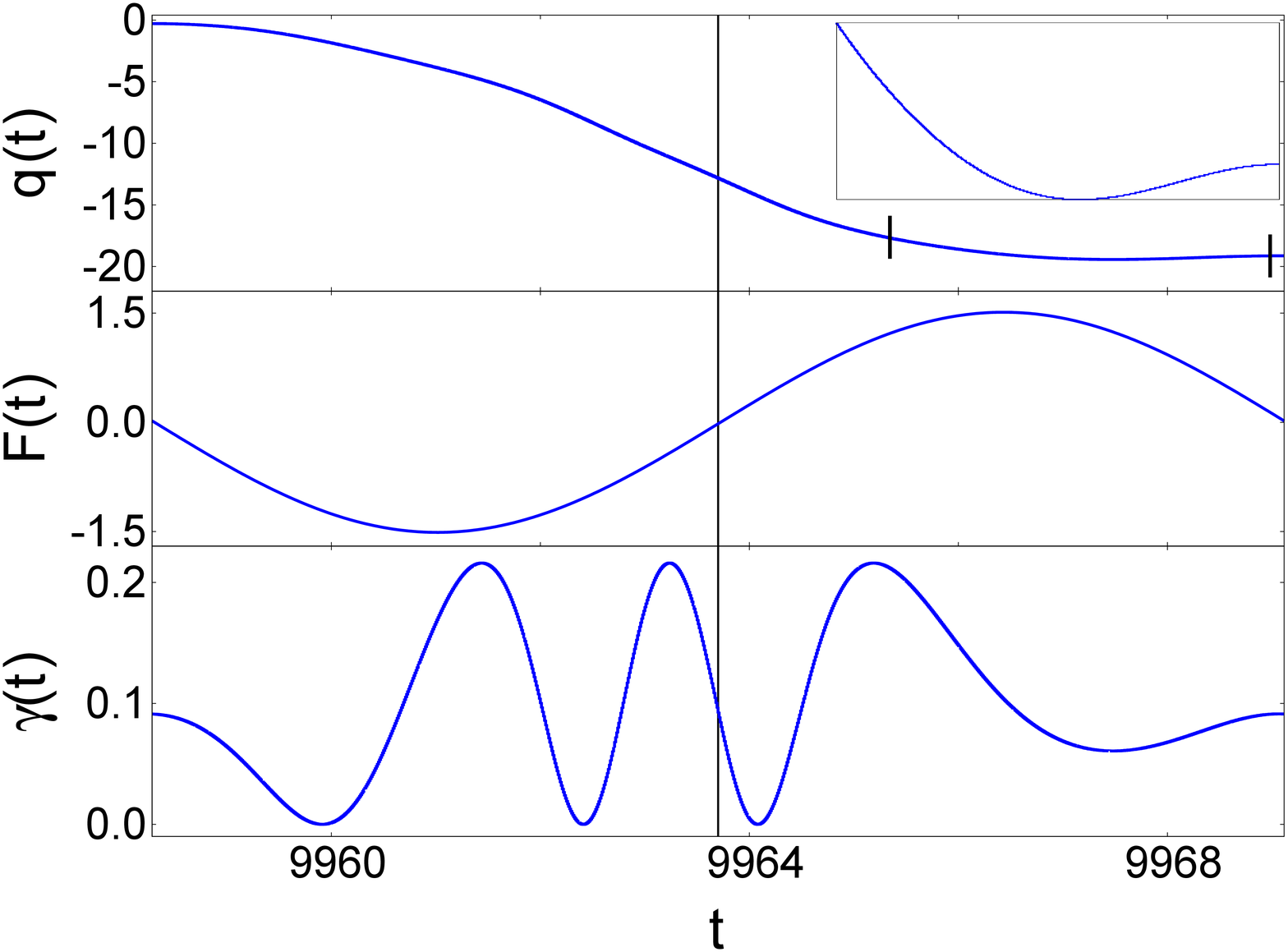}
\caption{The constituent parts of a solution exhibiting negative mobility over the course of a single period
of the external driving for $F=0.1$, $\lambda=1.0$, and the remaining parameters as in Fig.~\ref{fig:current}.
The top panel shows the evolution of the coordinate $q$, the middle panel shows the
time periodic driving, and the bottom panel shows the space-dependent damping. The vertical line in each
panel highlights the time when the driving changes from negative to positive.
Note that the coordinate $q$ in the top panel is shown mod$(8360)$. The inset
in the top panel is a magnification of the portion of the curve between the two
small vertical lines.}
\label{fig:anm}
\end{figure}

Now let's turn attention to the actual mechanism that allows a particle to run uphill. For simplicity, we
will look at a parameter set corresponding to the window of period one orbits seen in Fig.~\ref{fig:bifur}.
An example trajectory, with starting time $t\approx 9958.3$ coinciding with a
change from positive to negative external
driving, for this parameter set is given in Fig.~\ref{fig:anm} (top panel). When driving (middle panel)
becomes negative, the damping
strength (bottom panel) is approaching its minimum. Importantly, when the damping strength reaches its
minimum, the
driving strength is also close to its minimum of $-A$. This coordination between driving and damping allows
the
particle to run almost freely uphill. Subsequently, at $t \approx 9961$ the damping coefficient attains its
maximal value. However, the particle is being driven against the bias (by a driving force that is still close
to it minimum value), and continues to be so even after the damping coefficient has oscillated once more
between its minimum and maximum values.

As the driving becomes positive at $t \approx 9963.5$ (indicated by the vertical lines in the figure), the
damping coefficient is approaching its minimum. This results in a slowing down of the particle's ascent.
Eventually the particle's uphill motion ceases and it then follows the direction of the bias (see inset in
Fig.~\ref{fig:anm}). Importantly, this turning point occurs in the final stages (in the course of a single
period of the external driving) of positive driving, resulting in only short intervals of downhill motion.
This behaviour continues in a periodic fashion allowing the particle to travel large distances in the
direction opposite to the applied dc-bias force.

This mechanism, while sharing some of the characteristics of absolute negative mobility seen in previous
studies in that it depends on fine tuning of the external driving for said effect to occur, is unique as it
relies on the nonuniform damping to aid the uphill motion.

\begin{figure}
\includegraphics[height=4.8cm, width=5.8cm]{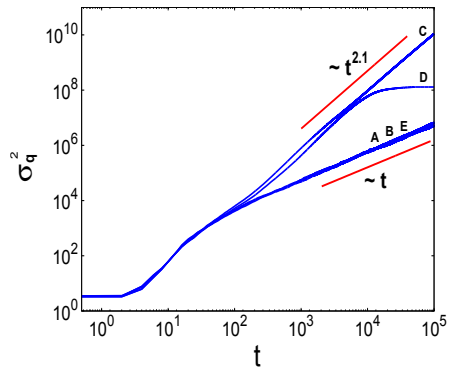}
\caption{The temporal evolution of the particle mean squared displacement for five values of the
inhomogeneity parameter $\lambda$: A) $\lambda=0$, B) $\lambda=0.2$, C) $\lambda=0.67$, D) $\lambda=1.0$, and
E) $\lambda=1.3$. The curves A,B, and E are only distinguishable upon magnification. The diffusive behaviour
is indicated by the (red) fitted lines.}
\label{fig:diffusion}
\end{figure}

To further characterise the motion we now look at the mean squared displacement for ensembles of particles,
i.e the rate of diffusion, given by

\begin{equation}
 \sigma_q^2(t) = \langle(q-\langle q \rangle)^2\rangle
\end{equation}

\noindent
where $\langle...\rangle$ indicates averages over ensemble. Typical \emph{normal} diffusion processes exhibit
a linear relationship with time, that is

\begin{equation}
 \sigma_q^2(t)\sim t^\alpha
\end{equation}

\noindent
with $\alpha=1$. However, with $\alpha \neq 1$ the diffusion becomes \emph{anomalous} -- either superdiffusive
($\alpha>1$), or subdiffusive ($\alpha<1$) \cite{Sancho1}. Fig.~\ref{fig:diffusion} shows the temporal
evolution of $\sigma_q^2(t)$ for five representative $\lambda$ values taken from
Fig.s~\ref{fig:current},\ref{fig:bifur}. These values are $\lambda=0$ (zero frictional inhomogeneity, chaotic
motion), $\lambda=0.2$ (nonzero frictional inhomogeneity, chaotic motion), $\lambda=0.67$ (regular coexisting
attractors, uphill motion), $\lambda=1$ (single periodic attractor, uphill motion), and $\lambda = 1.3$
(chaotic motion).

The early temporal evolution of $\sigma_q^2$, starting with every initial condition in the same potential
well, is similar in all
cases. Subsequently, at around $t\approx 10^2$ different regimes become apparent. The three $\lambda$ values
associated with chaotic (downhill) dynamics quickly settle into normal diffusive motion with an exponent
$\alpha \approx 1$
(lines A,B, and E in Fig.~\ref{fig:diffusion}). Interestingly, for the $\lambda$ values where ANM was
observed, the fitted exponent ($\alpha \approx 2.1$) shows that the motion is superdiffusive over a number of
decades. In fact, superdiffusion persists for the entire simulation in the case of $\lambda=0.67$. This is
not the case for $\lambda = 1$, where, after a number of decades, there is no diffusion. The reason for this
is that in phase space, when $\lambda=1$, only a single period 1 attractor exists, meaning that eventually all
initial
conditions evolve to this attractor. Thus, each trajectory undergoes the same motion resulting in the rate of
diffusion becoming zero. In contrast, for $\lambda=0.67$ there exists three attractors in phase space -- a
period 1 and a period 2 attractor exhibiting downhill motion, and a period 1 orbit exhibiting uphill motion.
These counterpropagating attractors yield the superdiffusive motion (remember diffusion here is ensemble
averaged).

\section{Summary}\label{sec:conclusion}
\noindent
We have studied the driven and damped dynamics of single particles evolving in a tilted periodic and
symmetric potential (the tilt being induced by a static dc-bias force). Unlike previous studies of such
systems where the damping coefficient remains constant, the system explored here contains a damping
coefficient
that is space-dependent. It has been shown that introducing a frictional inhomogeneity can result in some
interesting dynamics, most notably being the appearance of absolute negative mobility, i.e. solutions that run
against
an external load.

In more detail, with a zero dc-bias, a phase difference between the equally periodic potential and
nonuniform damping breaks a spatial symmetry of the system and allows for the emergence of a nonzero
current - as can be seen in Fig.~\ref{fig:current}. Increasing the (positive) dc-bias from zero has two
unexpected results. Firstly, the presence of the frictional inhomogeneity allows the particle to work against
a
significant load up to a critical value of the dc-bias $F=F_c$. Secondly, the current-response behaviour as a
function of the inhomogeneity parameter $\lambda$ and dc-bias value $F$ is quite remarkable. As $F$ is
increased from $F=0$ the window of $\lambda$ values exhibiting ANM, increases to almost three times its $F=0$
size, before shrinking to zero at $F=F_c$.

In addition, a heuristic explanation of the underlying mechanism producing such solutions
has revealed that the uphill motion relies on the space-dependent damping, and not just the frequency of the
driving, in contrast to previous studies. Further analysis has revealed that the uphill motion is
superdiffusive, at least on intermediate time periods, whereas the the downhill motion exhibits normal
diffusion.

\bibliography{References}

\end{document}